%% file: main.tex
\documentclass[journal]{IEEEtran}
 \usepackage{amsmath,amssymb}
 \usepackage{subfigure}
 \usepackage{graphicx,graphics,color,psfrag}
 \usepackage{cite,balance}
 \usepackage{algorithm}
 \usepackage{accents}
 \usepackage{amsthm}
 \usepackage{bm}
 \usepackage{url}
 \usepackage{algorithmic}
 \usepackage[english]{babel}
 \usepackage{multirow}
 \usepackage{enumerate}
 \usepackage{cases}
 \usepackage{stfloats}
 \usepackage{dsfont}
 \usepackage{color,soul}
 \usepackage{amsfonts}
 \usepackage{cite,graphicx,amsmath,amssymb}
 \usepackage{subfigure}
 \usepackage{fancyhdr}
 \usepackage{hhline}
 \usepackage{graphicx,graphics}
 \usepackage{array,color}
 \usepackage{amsmath}
 \usepackage{amsthm}
 \usepackage{framed} 
 \usepackage{multirow}
 \usepackage{hyperref}
 \usepackage{xcolor}
 \newif\ifreview
 \reviewfalse

 \ifreview
   \newcommand{\rev}[1]{\textcolor{blue}{#1}}
 \else
   \newcommand{\rev}[1]{#1}
 \fi

\include{header}

\DeclareMathSizes{10}{9.6}{5}{3}

\def\papertitle{ \huge The Universal Language of CSI: 
\\ Unifying Wireless Sensing Across Devices and Environments}

\IEEEoverridecommandlockouts 

\begin{document}

\title{ \fontsize{21}{21}\selectfont \papertitle}
\author{Jiayi Chen, Weiting Ou and Guangxu Zhu$^{*}$%
\thanks{This work was supported in part by Guangdong Major Project of  Basic and Applied Basic Research under Grant 2023B0303000001, in part by National Natural Science Foundation of China (Grant No. U25A20394, 62371313), in part by Guangdong Young Talent Research Project (Grant No. 2023TQ07A708), in part by the Shenzhen Science and Technology Program (Grant No. JCYJ20241202124934046), in part by by Shenzhen Loop Area Institute (Contract No. SLAI2026020007). \protect\\
\makebox[1.5em]{}J.~Chen and G.~Zhu are with the Shenzhen Research Institute of Big Data and The Chinese University of Hong Kong, Shenzhen, Guangdong, China. G.~Zhu is also with Shenzhen Loop Area Institute (SLAI), Shenzhen, Guangdong, China. W.~Ou is with the Shenzhen Research Institute of Big Data, Shenzhen, Guangdong, China. Emails: \{jiayichen5@link.cuhk.edu.cn, gxzhu@sribd.cn, ouweiting7@gmail.com\}. $^{*}$Corresponding authors: G.~Zhu.}%
}
\maketitle
\renewcommand{\headrulewidth}{0pt} 

\begin{abstract}
WiFi sensing based on Channel State Information (CSI) promises ubiquitous, device-free perception, yet current research remains trapped in a ``Tower of Babel''—fragmented into isolated silos where models are tailored to specific hardware ``dialects'', fixed environments, and narrow tasks. The primary bottleneck is the Heterogeneity Gap: the disparity in signal dimensions, sampling rates, and semantic labels that prevents cross-system understanding. To bridge this gap, we propose a foundation-model framework that treats CSI not merely as raw signals but as a structured language with a learnable universal grammar. We first curate and standardize a large collection of heterogeneous real-world CSI datasets, establishing a unified infrastructure that allows incompatible signal formats to be treated as a single corpus. Second, we introduce a modular architecture that acts as a universal translator where lightweight dataset-specific adapters ``tokenize'' diverse signal inputs into a shared latent vocabulary, while a shared self-supervised Transformer backbone learns the temporal syntax of human motion and environmental dynamics. This design decouples sensing ``semantics'' from hardware ``syntax''. Extensive evaluations show that by mastering this universal ``language'', our approach consistently outperforms task-specific baselines and exhibits strong generalization capability in new environments, achieving superior efficiency in few-shot scenarios. By effectively absorbing heterogeneity, the framework offers a path toward robust, general-purpose wireless sensing, mirroring the linguistic generalization observed in Large Language Models. The code implementation is available at: \url{https://github.com/cjychenjiayi/WiLLM}.
\end{abstract}


\IEEEpeerreviewmaketitle

\section{Introduction}
The evolution toward sixth-generation (6G) wireless networks is transforming communication systems from mere data delivery pipes into platforms that \emph{read} the physical world. This vision, referred to as \emph{Environmental Intelligence (EI)}, requires wireless networks not only to communicate but also to perceive their surroundings, interpret contextual dynamics, and interact with the physical environment. Achieving EI requires sensing capabilities deployable at scale across heterogeneous devices and environments, \rev{demanding that networks capture shared patterns underlying physical motion, which can be interpreted as a form of latent ``language''} \cite{ChannelGPT2025,WirelessGPT2025,AI2MMUM2025,MUSEFM2026}.

Wireless sensing offers a promising pathway toward this goal by leveraging existing communication infrastructure without requiring additional sensors. WiFi-based sensing has gained prominence due to its ubiquity, low deployment cost, and strong penetration in indoor environments. By exploiting Channel State Information (CSI), WiFi sensing systems capture fine-grained channel variations induced by human motion and environmental changes, enabling applications such as activity recognition, gesture sensing, indoor localization, and health monitoring \cite{Ma2018SignFi,Alsaify2020WiFiDataset,Wang2024XRF55}. However, despite steady progress, practical WiFi sensing remains trapped in a ``Tower of Babel'', i.e., models for specific hardware configurations or environments often become incoherent when transferred to others.

The fundamental limitation to scalability is not a lack of model sophistication, but the pervasive heterogeneity of CSI itself---a challenge we frame as the \rev{lack of a shared representation that can generalize across heterogeneous settings}. In practice, CSI measurements are inseparably entangled with the specific ``dialects'' imposed by hardware platforms, signal acquisition pipelines, and physical environments. As a result, CSI datasets collected by different research efforts differ in signal representations, sampling rates, temporal characteristics, and labeling conventions \cite{Alsaify2020WiFiDataset,Wang2024XRF55,Zhang2022Widar30}. Even for sensing tasks with identical high-level descriptions, the underlying signal distributions may vary significantly across datasets such that they resemble entirely different languages. We refer to this systemic challenge as the \emph{CSI Heterogeneity Gap}.

\rev{At a system level, the CSI heterogeneity gap arises from the entanglement of three complementary ``linguistic'' dimensions. \emph{System heterogeneity} stems from differences in hardware platforms and signal acquisition pipelines, analogous to variations in accents or phonetics. \emph{Scenario heterogeneity} reflects environment-dependent propagation effects, corresponding to contextual variation in signal transmission. \emph{Semantic heterogeneity} arises from inconsistent sensing objectives, label definitions, and annotation protocols, akin to mismatched vocabularies. Together, these dimensions capture distinct sources of variation from hardware acquisition, physical environments, and task semantics.} Together, these factors induce distribution shifts that pose a fundamental obstacle to unified modeling and cross-dataset generalization \cite{Zhang2022Widar30,wang2026wifisurvey}. Most existing WiFi sensing approaches address this challenge through task-specific designs \cite{Ma2018SignFi,Ding2020RFNet}, effectively training models ``fluent'' in a single narrow ``dialect''. While such approaches achieve strong performance under controlled conditions, they remain incompatible with multi-dataset training and generalize poorly across real-world deployments, hindering reproducibility and the development of \rev{more generalizable} sensing systems \cite{wang2026wifisurvey}.

Meanwhile, foundation models have transformed domains such as natural language processing and computer vision by learning transferable representations from large heterogeneous data. Just as large language models (LLMs) capture the underlying structure of language across topics, styles, and domains, wireless sensing should move toward learning shared structures in CSI dynamics. \rev{In this context, CSI measurements can be viewed as signal-frame sequences whose temporal dependencies reflect motion and multipath propagation, analogous to token relationships in language.} Existing wireless foundation models mainly focus on communication-centric tasks such as channel modeling and protocol intelligence \cite{ChannelGPT2025,WirelessGPT2025}, or employ LLMs only for high-level task inference without unified low-level signal representation learning \cite{Ren2025WiChat,LLM4CP2024}. Consequently, learning shared representations across heterogeneous CSI sources remains largely unexplored.

\begin{figure*}[t]
    \centering
    \scalebox{1.1}[1.00]{%
        \includegraphics[width=0.8\textwidth]{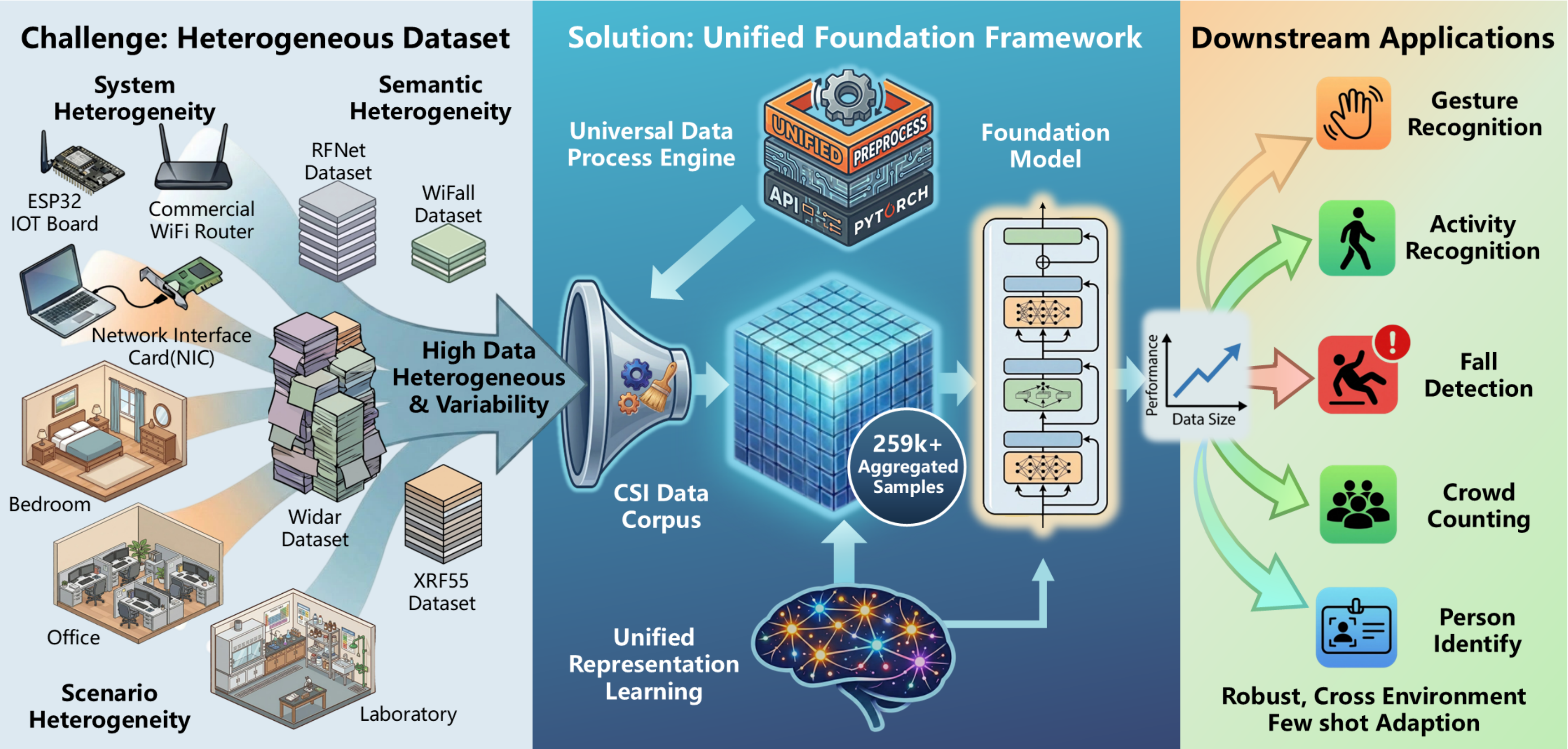}
    }
    \caption{Overview of the proposed unified WiFi CSI sensing framework for heterogeneous datasets.}
    \label{fig:design}
    \vspace{-2.5mm}
\end{figure*}

Motivated by this gap, we propose a paradigm shift from task-specific models toward a \emph{CSI Foundation Model} that treats heterogeneity as a first-class design constraint. As illustrated in Fig.~\ref{fig:design}, the proposed framework acts as a \rev{unified translation mechanism} for WiFi sensing. Instead of rigid signal-level alignment, dataset-specific adapters \rev{map heterogeneous CSI inputs into latent token sequences through lightweight embedding modules}, while a Transformer backbone learns the temporal structure of wireless channel dynamics. By decoupling the ``syntax'' of physical motion from the ``dialects'' of hardware and environments, this framework moves toward \rev{a unified framework for more generalizable WiFi CSI sensing}.

\section{Related Work and Motivation}
To establish a \rev{shared representation framework for wireless sensing}, it is first necessary to understand why the current research landscape remains deeply fragmented and how the paradigm of foundation models offers a viable path forward. This section deconstructs the ``Tower of Babel'' in existing CSI-based sensing research and motivates the shift toward a unified, linguistic approach to wireless signal processing.

\subsection{The Tower of Babel: Deconstructing CSI Heterogeneity}

Channel State Information (CSI) represents the fundamental ``speech'' of the wireless channel, capturing how radio waves propagate from transmitter to receiver. For a subcarrier $k$, the CSI $H(k)$ is a complex-valued quantity that jointly encodes amplitude attenuation and phase shift. While CSI theoretically contains rich semantic information about the physical environment, such as human presence and motion, in practice this information is heavily entangled with the specific ``dialect'' of the sensing hardware and deployment context.

We categorize this fragmentation, referred to as the \emph{CSI Heterogeneity Gap}, into three linguistic barriers:

\begin{enumerate}
    \item \textbf{System Heterogeneity (the ``Accent'').}
    Just as the same word sounds different in different accents, identical physical motions can produce vastly different CSI signatures across hardware platforms. Variations in center frequency (e.g., 2.4~GHz versus 5~GHz), bandwidth, antenna topology, and radio front-end fundamentally alter CSI structure. Moreover, hardware imperfections such as carrier frequency offset (CFO), sampling rate offset (SRO), and automatic gain control (AGC) introduce platform-specific distortions difficult to normalize across devices \cite{Ding2020RFNet,Zhang2022Widar30}. As a result, many existing models implicitly overfit to these hardware ``accents'' and fail to recognize the same underlying motion when observed through a new device.

    \item \textbf{Scenario Heterogeneity (the ``Context'').}
    Wireless propagation is inherently contextual. Multipath effects caused by reflections from walls, furniture, and objects mean that the ``grammar'' of signal propagation varies significantly across environments. A gesture performed in a laboratory may exhibit different spectral and temporal signatures when performed in a living room or hallway. Existing approaches often treat environmental characteristics as fixed, producing models that generalize poorly beyond their training environments \cite{Zhang2022Widar30,wang2026wifisurvey}.

    \item \textbf{Semantic Heterogeneity (the ``Vocabulary'').}
    There is no standardized dictionary for wireless sensing. Different datasets adopt incompatible labeling conventions, sensing objectives, and annotation protocols, and often operate at different sampling rates and temporal granularities. \rev{This lack of a shared vocabulary prevents aggregating diverse datasets into a coherent corpus for learning transferable sensing representations.} \cite{Alsaify2020WiFiDataset,wang2026wifisurvey}.
\end{enumerate}

\rev{These three types capture distinct sources of variation: system heterogeneity originates from hardware and signal acquisition, scenario heterogeneity reflects environment-dependent propagation effects, and semantic heterogeneity arises from task definitions and labeling protocols.}

\subsection{Foundation Models: Toward Shared CSI Structure}

A similar fragmentation once characterized natural language processing (NLP), where early systems struggled to generalize across domains, styles, and vocabularies. The emergence of foundation models such as BERT and GPT demonstrated that training on large and diverse corpora allows models to learn the statistical structure of language—its syntax and grammar—enabling robust generalization to unseen tasks and domains. We argue that CSI exhibits a comparable latent structure. \rev{In particular, CSI measurements can be viewed as sequences of signal frames, where each frame plays a role analogous to a token, and their temporal dependencies reflect underlying physical processes such as motion and multipath propagation.} This analogy motivates a foundation-model-style approach to wireless sensing:

\begin{itemize}

\item \textbf{CSI Corpus Construction.}
Foundation models in NLP benefit from aggregating heterogeneous text into a single corpus, allowing models to learn general linguistic structures. In contrast, wireless sensing datasets remain fragmented due to differences in hardware platforms, environments, signal acquisition pipelines, and labeling protocols. Constructing a unified CSI corpus across datasets is therefore essential for large-scale representation learning and for revealing shared statistical structures in heterogeneous CSI measurements.

\item \textbf{CSI-Native Tokenization.}
In NLP, raw text is decomposed into tokens that abstract surface variations and provide standardized representations for downstream models. Similarly, raw CSI frames can be transformed into \rev{latent token sequences} that decouple sensing-relevant dynamics from hardware-specific signal dimensions. Since CSI datasets differ in subcarrier configurations, antenna layouts, sampling rates, and signal formats, CSI-aware tokenization is required to project heterogeneous inputs into a shared representation space. In our framework, dataset-specific embedding modules serve as tokenizers that map raw CSI sequences into \rev{continuous token representations in a shared embedding space}.

\item \textbf{CSI Representation Learning.}
Transformer architectures capture long-range dependencies between tokens through self-attention, enabling contextual understanding of language. Similarly, wireless sensing signals contain temporal patterns and spatial correlations across antennas and subcarriers due to multipath propagation. By projecting heterogeneous CSI inputs into a shared embedding space and employing a shared Transformer backbone, the model learns global contextual relationships within CSI sequences and captures the spatial--temporal structure of wireless channel dynamics.

\end{itemize}

Recent efforts have explored foundation models for wireless systems \cite{ChannelGPT2025,WirelessGPT2025}, but these works mainly focus on communication-centric tasks such as channel modeling, beamforming, or protocol intelligence. From a linguistic perspective, they function more like ``spell-checkers'' for transmission rather than ``translators'' for sensing, as they do not address the semantic challenge of interpreting human behavior across heterogeneous devices and environments. \rev{Although both sensing and communication tasks are driven by CSI, their representation requirements differ: sensing benefits from invariance to device-specific distortions and emphasizes motion-induced temporal variations, whereas communication tasks such as channel prediction, beam management, CSI feedback compression, and link adaptation often require preserving fine-grained channel-state and link-quality information. Extending the current sensing-oriented framework to these tasks would therefore require communication-aware heads and multi-objective losses, which we leave as future work toward ISAC-oriented CSI foundation model}

To address this limitation, we introduce a unified CSI foundation modeling framework. Section~\ref{sec:unified_dataset} presents the construction of a unified CSI corpus from heterogeneous datasets. Section~\ref{sec:tokenization} introduces CSI-aware tokenization that maps raw CSI signals into latent tokens, and Section~\ref{sec:temporal_modeling} describes a shared Transformer-based temporal modeling framework for robust motion understanding across diverse sensing scenarios.

\begin{table*}[t]
\centering
\caption{Summary of public WiFi CSI datasets}
\label{tab:dataset_summary}
\setlength{\tabcolsep}{5pt}
\renewcommand{\arraystretch}{1}
\begin{tabular}{|l|c|c|c|c|c|c|c|}
\hline
\textbf{Dataset} & \textbf{Samples} & \textbf{Classes} & \textbf{Subcarriers} & \textbf{Timestamps} & \textbf{Frequency} & \textbf{Samples Per Class} & \textbf{Task Type} \\
\hline
RFNet~\cite{Ding2020RFNet}            
& 9,600  & 6   & 60  & 512  & --        & 1,600 & Gesture Recognition \\
\hline
SignFi~\cite{Ma2018SignFi}           
& 3,200  & 50  & 30  & 256  & 5\,GHz    & 64    & Sign Language Recognition \\
\hline
CSLOS~\cite{Alsaify2020WiFiDataset}                                 
& 6,760  & 15  & 90  & 512  & 2.4\,GHz & 450.7 & Activity Recognition \\
\hline
XRF55~\cite{Wang2024XRF55}            
& 17,600 & 55  & 270 & 1000 & 5\,GHz    & 320   & Activity Recognition \\
\hline
NTUHumanID~\cite{Yang2022EfficientFi} 
& 294    & 14  & 114 & 2000 & 5\,GHz    & 21.0  & Human Identification \\
\hline
NTUHAR~\cite{Yang2022EfficientFi}     
& 936    & 6   & 114 & 2000 & 5\,GHz    & 156.0 & Activity Recognition \\
\hline
Widar3.0~\cite{Zhang2022Widar30}      
& 213,614 & 22 & 90  & 100  & 5.825\,GHz & 9,709.7 & Activity Recognition \\
\hline
WiCount (ours)                           
& 624    & 4   & 52  & 100  & 2.4\,GHz        & 156.0 & Crowd Counting \\
\hline
WiFallAct (ours)                         
& 1,409  & 5   & 52  & 100  & 2.4\,GHz        & 281.8 & Fall Detection \\
\hline
WiGestureAct (ours)                     
& 2,214  & 6   & 52  & 100  & 2.4\,GHz        & 369.0 & Gesture Recognition \\
\hline
WiFallID (ours)                  
& 1,409  & 10  & 52  & 100  & 2.4\,GHz        & 140.9 & Human Identification \\
\hline
WiGestureID (ours)                      
& 2,214  & 8   & 52  & 100  & 2.4\,GHz        & 276.8 & Human Identification \\
\hline
\textbf{Total / Range}
& \textbf{259,774}
& \textbf{4--55}
& \textbf{30--270}
& \textbf{100--2000}
& \textbf{--}
& --
& \textbf{6 Task Categories} \\
\hline
\end{tabular}
\vspace{-1mm}
\end{table*}

\section{CSI Corpus Construction}
\label{sec:unified_dataset}
A critical bottleneck in scalable WiFi CSI sensing lies in data infrastructure rather than model design. Public CSI datasets are tightly coupled with specific devices, environments, and acquisition pipelines, and thus remain fragmented, heterogeneous, and difficult to reuse across studies. Consequently, most WiFi sensing works still rely on a single dataset collected under fixed settings, which limits cross-dataset learning, systematic generalization analysis, and reproducibility.

While recent efforts have surveyed publicly available WiFi CSI sensing datasets (e.g.,~\cite{wang2026wifisurvey,MUSEFM2026}), they largely provide taxonomy-level summaries and dataset lists, leaving data formats, preprocessing pipelines, and code interfaces inconsistent and impractical for joint training. \textbf{\rev{To address this limitation, our framework consolidates heterogeneous CSI datasets into a unified corpus that explicitly supports cross-dataset representation learning, and provides a PyTorch-ready interface for consistent loading and preprocessing across datasets.}} Rather than processing each dataset in isolation, we curate and operationalize a diverse collection of real-world CSI datasets from public sources~\cite{wang2026wifisurvey, Alikhani2025LWM}, enabling reproducible multi-source training and evaluation for WiFi sensing.

\subsection{Dataset Curation and Diversity}

We curate WiFi CSI datasets spanning a broad range of indoor sensing tasks, including gesture recognition, human activity recognition, human identification, fall detection, and crowd counting. The corpus includes both publicly available datasets and datasets collected using in-house experimental setups. In this work, we activate a representative subset of public datasets for model training and evaluation, including RFNet~\cite{Ding2020RFNet}, XRF55~\cite{Wang2024XRF55}, CSLOS~\cite{Alsaify2020WiFiDataset}, NTUHumanID~\cite{Yang2022EfficientFi}, NTUHAR~\cite{Yang2022EfficientFi}, and Widar3.0~\cite{Zhang2022Widar30}. Beyond these public datasets, the corpus also includes several in-house CSI datasets developed within our research group, including \emph{WiCount}, \emph{WiFallAct}, \emph{WiGestureAct}, \emph{WiFallID}, and \emph{WiGestureID}. While fully supported by the unified dataset interface, these datasets are internal resources rather than contributions of this paper.\footnote{The framework already integrates more datasets, but only a representative subset is activated for training and evaluation in this work. The rest are included in the unified interface but not used in this paper's experiments.}

Importantly, even within the same sensing category, different datasets define distinct class semantics, motion patterns, sensing objectives, and experimental protocols. \rev{Rather than strictly normalizing datasets, the curated corpus preserves diversity, exposing the model to signal variations across heterogeneous devices and environments.} This design reflects the heterogeneity that \rev{more generalizable} WiFi sensing systems must handle in real deployments \cite{wang2026wifisurvey,Zhang2022Widar30}. All datasets are based on real CSI measurements collected using diverse WiFi devices and setups, and vary substantially in subcarrier layouts, antenna configurations, sampling rates, sequence lengths, and data formats. Table~\ref{tab:dataset_summary} summarizes the key characteristics of the curated datasets. Together, these datasets form a unified corpus within our data infrastructure, enabling consistent access and joint learning across heterogeneous WiFi sensing scenarios.

\subsection{Unified Dataset Interface for WiFi Sensing}

To transform fragmented datasets into a reusable system resource, we design a unified dataset interface that standardizes how heterogeneous CSI samples and labels are accessed and processed. A key design choice is to avoid rigid signal-level alignment across datasets, which may distort CSI measurements or remove dataset-specific channel characteristics, as highlighted by prior studies on cross-domain WiFi sensing and generalization challenges. Instead, we adopt a dataset-aware preprocessing strategy where anomaly removal, phase sanitization, amplitude normalization, and noise suppression are performed locally with dataset-specific configurations. \rev{This design preserves dataset-specific characteristics while reducing measurement artifacts, thereby enabling more robust cross-dataset learning.}

After preprocessing, each dataset is wrapped into a common Python-based abstraction that exposes consistent input--output behavior regardless of file format, signal dimension, antenna configuration, or task definition. Through this interface, heterogeneous CSI datasets can be instantiated via a unified configuration and directly integrated into a standard PyTorch \texttt{DataLoader} for multi-source sampling and batching, enabling cross-dataset joint training and evaluation. \rev{By decoupling dataset-specific engineering from model development, this interface enables scalable deployment across sensing platforms and improves reproducibility in large-scale WiFi sensing systems.} \cite{wang2026wifisurvey}. The complete dataset corpus, preprocessing pipeline, and interface implementation
are publicly available at \url{https://github.com/cjychenjiayi/WiLLM}.

\begin{figure*}[t]
    \centering
    \scalebox{1.04}[1.00]{%
        \includegraphics[width=0.85\textwidth]{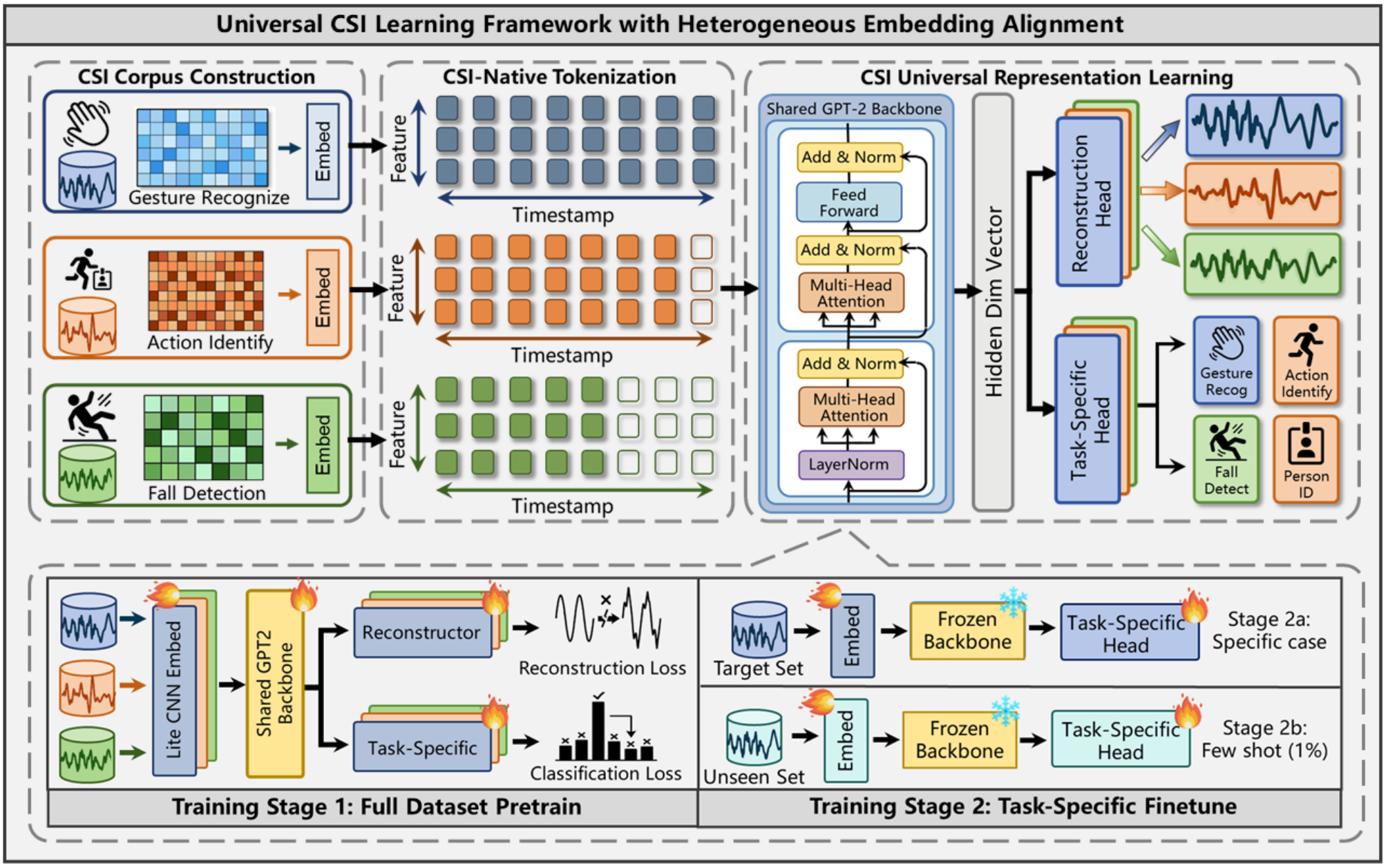}
    }
    \caption{A unified pipeline for adapting heterogeneous WiFi CSI datasets to a shared foundation-model training framework.}
    \label{fig:strategy}
    \vspace{-1mm}
\end{figure*}

\section{A Universal CSI Learning Framework with Heterogeneous Embedding Alignment}
\label{sec:universal_framework}

Given the heterogeneous CSI corpus constructed in the previous section, the next challenge is to learn a unified representation that can generalize across datasets with fundamentally different sensing pipelines. Directly enforcing alignment at the raw signal level is undesirable, as CSI measurements collected under different hardware configurations and preprocessing pipelines may contain incompatible physical structures. \rev{Moreover, manually designing a fully unified preprocessing pipeline across datasets would introduce many dataset-dependent hyperparameters, making the framework difficult to scale and less amenable to end-to-end optimization.} Inspired by heterogeneous pre-trained Transformer architectures that align representations in embedding space rather than input space~\cite{Wang2024HPT, Alikhani2025LWM}, we adopt the same principle for CSI modeling. As illustrated in Fig.~\ref{fig:strategy}, heterogeneous CSI sequences are projected into a shared latent representation space, enabling joint modeling across datasets. \rev{In this way, dataset-specific structural differences are absorbed by lightweight front-end projections, while the shared backbone focuses on learning temporal patterns that are transferable across sensing settings.} This design decouples dataset heterogeneity from \rev{shared} representation learning and supports scalable cross-dataset training.

\subsection{CSI-Native Tokenization via Convolutional Embedding}
\label{sec:tokenization}

CSI datasets exhibit heterogeneous structural characteristics. In some scenarios, temporal dynamics dominate while subcarrier structure remains relatively stable, resembling sequential time-series data. In others, strong correlations exist across antennas and adjacent subcarriers, forming structured frequency-domain patterns. This mixed temporal--spectral nature makes purely sequence-based modeling insufficient, while directly treating CSI matrices as images may overlook their propagation-aware structure.

\rev{To absorb dataset-level heterogeneity, we employ dataset-specific lightweight one-dimensional CNNs as front-end tokenizers. Each CNN projects native CSI amplitude sequences into a shared latent space while preserving signal structures. Operating along the temporal axis, these modules capture short-range temporal dependencies and local subcarrier/antenna correlations induced by multipath propagation, while mitigating hardware-specific noise and preprocessing variations. In this way, heterogeneous CSI measurements are converted into latent token sequences with a unified feature dimension for subsequent shared-backbone modeling.}

\subsection{CSI Universal Representation Learning}
\label{sec:temporal_modeling}

The latent token sequences produced by the embedding modules are processed by a shared Transformer backbone for global representation learning. The backbone follows a GPT-2 style architecture and is trained from scratch on the heterogeneous CSI corpus. By sharing the same backbone parameters across datasets, the model accumulates representation learning capacity from diverse sensing systems and learns transferable spatial--temporal patterns of wireless channel dynamics. \rev{Once heterogeneous CSI measurements are projected into a common embedding space, the shared backbone focuses less on dataset-specific formats and more on recurring temporal dynamics across devices, environments, and tasks.} Through stacked self-attention layers, the backbone captures long-range dependencies in CSI sequences while promoting invariance to hardware platforms and deployment environments.

Training follows a two-stage strategy. In the first stage, the shared backbone is jointly trained across datasets, while dataset-specific heads are attached for signal reconstruction. These heads are not shared and map the backbone outputs back to the original CSI signal space, producing a reconstruction loss for each dataset. \rev{The reconstruction target is the original CSI amplitude sequence, serving as an information-preserving constraint that helps the shared backbone retain signal-relevant features and reduce projection-induced information loss.} To mitigate bias toward datasets with larger training volumes, dataset sampling weights are set to be inversely related to the \rev{square root of the average number of samples per class}, whose statistics are summarized in Table~\ref{tab:dataset_summary}. \rev{This weighting strategy helps balance the influence of different datasets during joint training and compensates for disparities in data scale and task difficulty.} In the second stage, the pretrained backbone is adapted to downstream sensing tasks through lightweight task heads optimized with regression or prediction losses. This combination of shared representation learning and dataset-specific lightweight heads enables robust WiFi sensing under both full-supervision and few-shot settings.

\begin{table*}[htbp]
\centering
\caption{Overall classification accuracy across heterogeneous WiFi CSI datasets}
\label{tab:overall_compare}
\setlength{\tabcolsep}{5pt}
\renewcommand{\arraystretch}{1}
\begin{tabular}{|l|c|c|c|c|c|c|c|c|c|c|c|}
\hline
\textbf{Method} 
& \textbf{RFNet} 
& \textbf{XRF55} 
& \textbf{NTUID} 
& \textbf{NTUHAR} 
& \textbf{WiCount} 
& \textbf{WiFallAct} 
& \textbf{WiGestAct} 
& \textbf{WiFallID} 
& \textbf{WiGestID} 
& \textbf{WIDAR3}
& \textbf{Avg} \\
\hline
MLP        
& 84.75 
& 82.70 
& 80.29 
& 96.59 
& \underline{87.18} 
& 54.78 
& 95.14 
& 67.70 
& \underline{99.10} 
& 45.15
& 79.34 \\
\hline
ResNet50   
& 79.32 
& \textbf{96.72} 
& 74.82 
& \underline{98.12} 
& 71.58 
& 31.46 
& 47.84 
& 52.25 
& 81.29 
& \underline{75.33}
& 70.87 \\
\hline
ResNet101  
& 76.40 
& \underline{95.72} 
& 78.43 
& 90.62 
& 58.26 
& 32.20 
& 46.40 
& 51.69 
& 76.08 
& \textbf{75.62}
& 68.14 \\
\hline
RNN        
& 73.78 
& 67.35 
& 77.39 
& 93.13 
& 83.97 
& 44.38 
& 92.09 
& 66.29 
& 98.56 
& 13.73
& 71.07 \\
\hline
GRU        
& 57.16 
& 88.47 
& 91.83 
& \textbf{99.38} 
& 85.90 
& 55.06 
& 94.60 
& 69.10 
& 98.38 
& 68.24
& 80.81 \\
\hline
LSTM       
& 73.47 
& 33.29 
& 88.12 
& 96.88 
& \underline{87.18} 
& 51.97 
& 93.38 
& 68.54 
& 98.74 
& 65.40
& 75.70 \\
\hline
BiLSTM     
& 71.65 
& 42.50 
& 87.35 
& 95.00 
& 83.33 
& 53.09 
& 95.68 
& 67.42 
& \underline{99.10} 
& 67.00
& 76.21 \\
\hline
Transformer
& 64.64 
& 91.67 
& \underline{92.76} 
& \underline{98.12} 
& 81.92 
& 39.89 
& 91.01 
& 62.36 
& 98.02 
& 69.68
& 79.01 \\
\hline
Pretrain (ours)
& \underline{95.96} 
& 88.07 
& 91.21 
& 88.64 
& 84.62 
& \underline{77.90} 
& \underline{97.83} 
& \underline{87.25} 
& \textbf{100.00} 
& 70.34
& \underline{88.18} \\
\hline
Finetune (ours)
& \textbf{97.96} 
& 90.59 
& \textbf{93.77} 
& 94.32 
& \textbf{89.10} 
& \textbf{84.70} 
& \textbf{98.92} 
& \textbf{90.09} 
& \textbf{100.00} 
& 71.18
& \textbf{91.06} \\
\hline
\end{tabular}
\vspace{-2mm}
\end{table*}

\section{Experimental Evaluation}

This section evaluates the effectiveness and robustness of the proposed shared CSI representation framework across heterogeneous WiFi sensing datasets, with an emphasis on cross-dataset consistency and generalization rather than per-dataset performance tuning. A unified training and evaluation protocol is adopted throughout all experiments to ensure that observed performance differences primarily reflect representation quality under diverse environments, hardware platforms, and signal formats. Both full-supervision and limited-label scenarios are considered to assess scalability and sample efficiency in realistic deployment settings. 

\subsection{Experimental Setup}

All experiments follow a unified multi-dataset training protocol to evaluate robustness across heterogeneous CSI datasets. The model adopts a shared GPT-2 style Transformer backbone to capture long-range temporal dependencies in CSI sequences. The backbone architecture is used without pretrained weights and is trained entirely from scratch. Raw CSI inputs are first processed by a lightweight temporal embedding module consisting of two 1D convolutional layers, producing token features with a dimension of 256. The maximum sequence length is set to 1024 timestamps, where longer sequences are cropped during training.

On top of the shared backbone, lightweight dataset-specific heads are attached for downstream learning. Two types of heads are used: a reconstruction head that maps backbone outputs back to the original CSI signal space, and a task head that predicts sensing outputs. Both heads are implemented using simple multilayer perceptrons and are not shared across datasets. Training jointly optimizes a reconstruction loss and a task-level prediction loss. To mitigate bias toward datasets with larger training volumes, dataset sampling weights are set to be inversely related to the average number of samples per class, whose statistics are summarized in Table~\ref{tab:dataset_summary}. All models are implemented in PyTorch and optimized using Adam. The batch size is set to 256, and all experiments are conducted under identical training settings across datasets.

\subsection{Performance Across Heterogeneous Datasets}

\begin{figure}[htbp]
\centering
\includegraphics[width=0.9\linewidth]{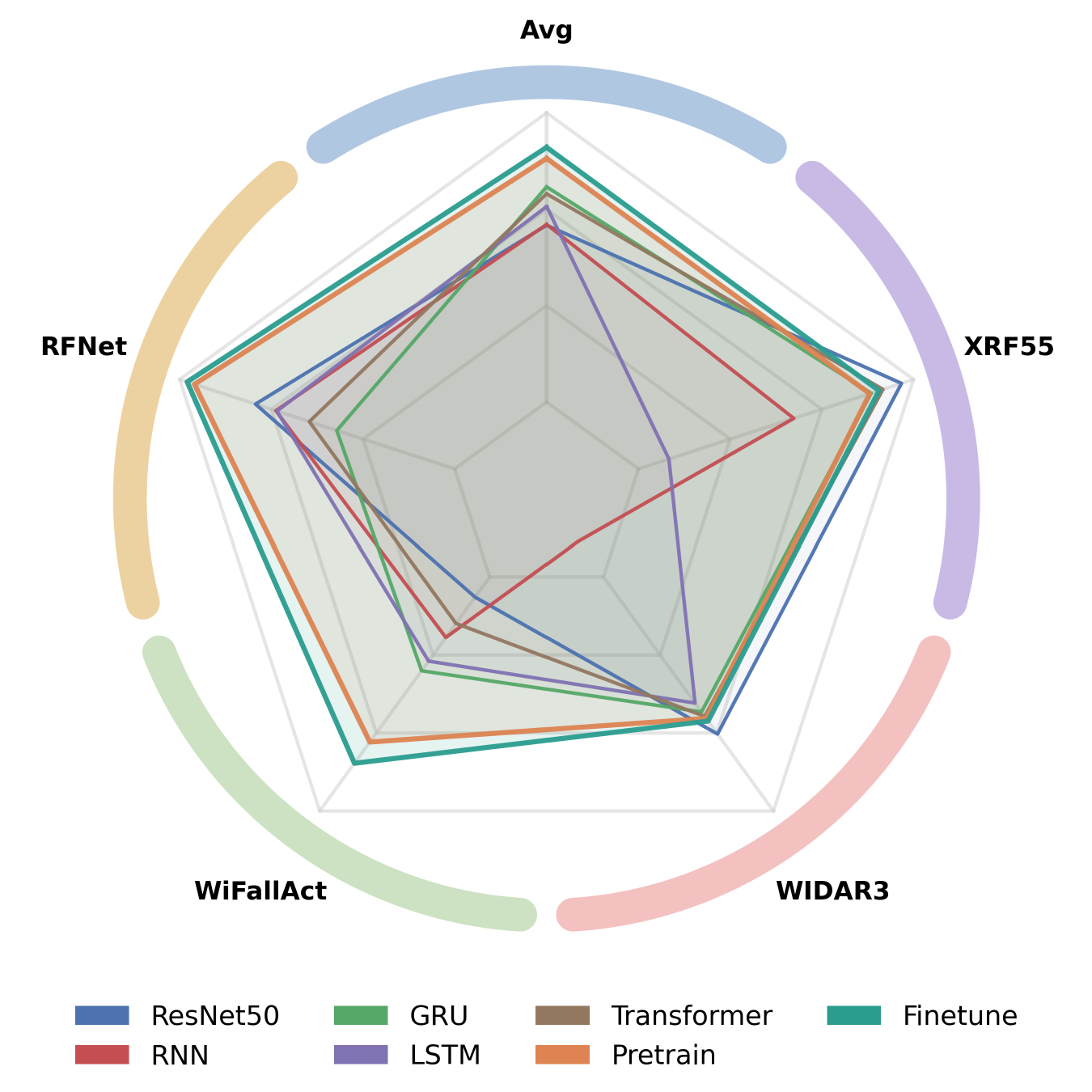}
\caption{Cross-dataset performance comparison under full supervision.}
\label{fig:alldata}
\end{figure}

We evaluate all models under full supervision, where all labeled samples in each dataset are used for training. The quantitative accuracy comparison across WiFi CSI datasets is summarized in Table~\ref{tab:overall_compare}, while Fig.~\ref{fig:alldata} provides a visual comparison of model behavior across representative datasets. As shown in Table~\ref{tab:overall_compare}, conventional task-specific models exhibit large performance variations across datasets. CNN-based architectures and recurrent networks achieve strong accuracy on certain benchmarks, yet experience substantial degradation on others, indicating strong dependence on signal formats, environments, and sensing scenarios. This inconsistency reflects the difficulty of directly transferring models optimized for a single dataset to heterogeneous CSI conditions.

In contrast, the proposed pretrained and finetuned models consistently outperform baseline methods in terms of average accuracy and maintain stable performance across all evaluated datasets. The balanced performance profile in Fig.~\ref{fig:alldata} further illustrates that the shared backbone avoids severe accuracy drops on challenging scenarios, demonstrating strong robustness to CSI heterogeneity. \rev{Importantly, this improvement comes from joint training across heterogeneous datasets rather than per-dataset tuning, encouraging the backbone to capture recurring temporal structures instead of dataset-specific formats.} Overall, these results suggest that the proposed framework effectively captures \rev{shared} temporal patterns in CSI signals and enables reliable performance across diverse WiFi sensing tasks without dataset-specific model redesign.

\subsection{Robustness Under Limited Supervision}

\begin{figure}[!t]
\centering
\scalebox{1.00}[1.00]{%
    \includegraphics[width=1\linewidth]{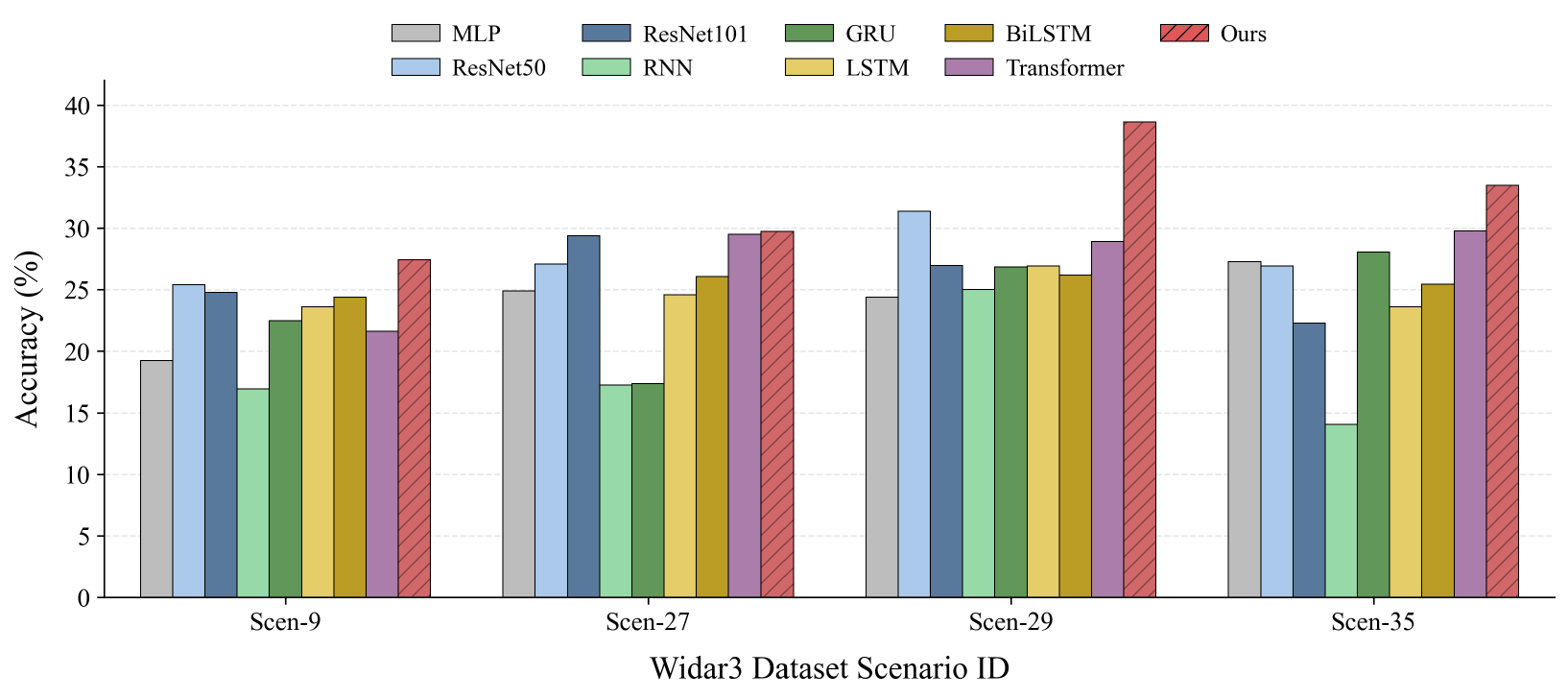}
}
\caption{Few-shot performance across different scenarios.}
\label{fig:fewshot}
\vspace{-1mm}
\end{figure}

To evaluate model robustness in a realistic data setting, we further conduct few-shot experiments on the Widar3 dataset. Following a same-scenario protocol, only 1\% of labeled samples in each scenario are used for training, while the remaining data are reserved for evaluation. The performance across representative scenarios is reported in Fig.~\ref{fig:fewshot}. Under this extreme data scarcity, all task-specific baseline models experience substantial accuracy degradation. CNN-based architectures, recurrent networks, and the Transformer trained from scratch fail to maintain reliable performance, with large variations observed across different scenarios.

In contrast, the proposed pretrained backbone consistently achieves the highest accuracy across all evaluated scenarios and exhibits significantly improved stability under limited supervision. Even with only 1\% labeled data, the model maintains strong performance without severe degradation, indicating that pretraining enables effective transfer of \rev{shared and transferable} temporal structures in CSI signals. \rev{The additional gain achieved by the finetuned model further supports the role of task-specific adaptation in refining these pretrained representations for downstream objectives.} These results show that the learned representations are sample-efficient and robust to data scarcity, reducing the labeling effort required for practical WiFi sensing deployment.

\section{Conclusion and Future Work}

\rev{This work departs from the fragmented ``Tower of Babel'' that has constrained wireless sensing~\cite{wang2026wifisurvey,Zhang2022Widar30}. By reframing the Heterogeneity Gap as a representation problem, we establish a universal CSI language that maps heterogeneous signals into a shared embedding space through unified data abstraction and representation learning~\cite{ChannelGPT2025,MUSEFM2026}.}

By curating heterogeneous real-world CSI datasets within a standardized infrastructure~\cite{Wang2024XRF55,Alsaify2020WiFiDataset} and introducing a modular Transformer backbone, our framework shows that hardware variations---including center frequency, bandwidth, and sampling rate---can still preserve \rev{learnable motion structures across datasets}. By decoupling environmental ``syntax'' from device ``accents'', the proposed architecture \rev{enables unified modeling across otherwise incompatible systems}.

The implications of this paradigm shift are threefold:
\begin{enumerate}
    \item \textbf{Unification:} Heterogeneous CSI datasets can be consolidated into a shared training corpus, revealing \rev{common patterns} across diverse wireless sensing platforms.
    
    \item \textbf{Fluency:} The Transformer backbone captures long-range temporal dependencies and achieves superior performance over task-specific models that often overfit to narrow hardware configurations~\cite{Zhang2022Widar30,Ren2025WiChat}.
    
    \item \textbf{Generalization:} \rev{The proposed framework exhibits strong few-shot adaptation, enabling the pretrained model to adapt to new environments with limited labels and improving cross-scenario transferability.}
\end{enumerate}

Looking ahead to the era of 6G and ubiquitous environmental intelligence, the ability to interpret radio signals across hardware platforms becomes increasingly important~\cite{LLM4CP2024, AI2MMUM2025}. Future work will extend this representation framework to additional sensing modalities, including mmWave and vision-based signals, enabling richer multimodal environmental understanding. Furthermore, continual and adaptive learning mechanisms will be explored to support long-term deployment under evolving sensing conditions~\cite{AI2MMUM2025}. Ultimately, this work \rev{provides a step toward} a new generation of wireless systems that not only transmit information, but also perceive and interpret the physical world through shared representations.

\bibliographystyle{IEEEtran}
\bibliography{BibDesk_File}

\vspace{4mm}

\centerline{B\small{IOGRAPHIES}}
\vspace{2mm}

\noindent{\bf Jiayi Chen} is pursuing the Ph.D. degree at Chinese University of Hong Kong, Shenzhen. His research interests include large model in wireless sensing and robotics.

\noindent{\bf Weiting Ou}
received his M.S. degree from the University of Electronic Science and Technology of China. His research interests include wireless sensing and tensor decomposition.

\noindent{\bf Guangxu Zhu}
(M'14) received his Ph.D. degree in The University of Hong Kong. His research interests include edge intelligence, distributed learning and wireless sensing.

\newpage

\end{document}

%% file: header.tex
\def\({\left(}
\def\){\right)}

\def\b0{{\mathbf{0}}}

